\documentclass{PoS}
\usepackage{amsfonts}
\usepackage{amsmath}
\usepackage{subfigure}
\usepackage{tipa}

\graphicspath{{./figure/}}
\newcommand{\ol}{ \overline }
\newcommand{\olSF}[2]{ \overline{({#1} \otimes {#2})} }
\newcommand{\ololSF}[2]{ \overline{\overline{({#1} \otimes {#2})}}}
\newcommand{\wtd}[1] { \widetilde{#1} }

\title{Non-perturbative Renormalization of Bilinear Operators 
  with Improved Staggered Quarks}

\ShortTitle{NPR of Bilinear Operators}

\author{\speaker{Jangho Kim}, Jongjeong Kim, Weonjong Lee 
\\
Lattice Gauge Theory Research Center, CTP, and FPRD, \\
Department of Physics and Astronomy, \\
Seoul National University, Seoul, 151-747, South Korea\\
E-mail: \email{wlee@snu.ac.kr}}

\author{Boram Yoon \\
Los Alamos National Laboratory, \\
Theoretical Division T-2, MS B283, \\
Los Alamos, NM 87545, USA \\
E-mail: \email{boram@lanl.gov}}

\author{SWME Collaboration}

\abstract{
We present renormalization factors for the bilinear operators obtained
using the non-perturbative renormalization method (NPR) in the RI-MOM
scheme with improved staggered fermions on the MILC asqtad lattices
($N_f = 2 + 1$).
We use the MILC coarse ensembles with $20^3 \times 64$ geometry and
$am_{\ell}/am_s = 0.01/0.05$.
We obtain the wave function renormalization factor $Z_q$ from the
conserved vector current and the mass renormalization
factor $Z_m$ from the scalar bilinear operator.
We also present preliminary results of renormalization factors for
other bilinear operators.
}

\FullConference{31st International Symposium on Lattice Field Theory -
  LATTICE 2013\\ July 29 - August 3, 2013\\ Mainz, Germany}

\begin{document}
\section{Introduction}
Recent calculation of $B_K$ in Refs.~\cite{Bae:2011ff,Bae:2010ki}
shows that the dominant systematic error ($\approx 4.4\%$) comes from
the matching factor obtained using the one-loop perturbation theory.
Hence, it becomes essential to reduce this error as much as possible.
One possibility is to calculate the matching factor using the two-loop
perturbation theory, which will reduce the systematic error down to
the $\approx 0.9\%$ level.
Another possibility is to obtain the matching factor using the
non-perturbative renormalization method (NPR) with the RI-MOM
\cite{Aoki:2007xm} and RI-SMOM scheme \cite{Sturm:2009kb}, which is
very likely to reduce the systematic error down to the $\approx 2\%$
level.
Here, we present preliminary results of renormalization factors of
bilinear operators calculated using NPR in the RI-MOM scheme with
improved staggered fermions.

\section{Bilinear Operator Renormalization}
A bilinear operator of staggered fermions is defined as
\begin{align}
O^{f_1 f_2}_{i}(y) 
= \sum_{A B} \sum_{c_1 c_2} \ol{\chi}^{f_1}_{i; c_1}(y_A) 
\olSF{\gamma_{S}}{\xi_F}_{AB}[U_{i; AB}]_{c_1 c_2}(y)
\chi^{f_2}_{i; c_2}(y_B) \,,
\end{align}
where $i$ is a gauge configuration index. 
$c_i$ are color indices and $f_i$ flavor indices. 
The $y$ represents a coordinate of the hypercube with its lattice
spacing $2a$.
The indices $A$, $B$ are hypercubic vectors such as $A = (0,1,1,0)$
for example.
Here, we use the notation of $y_A = 2y + A$. 
$[U_{i;AB}]_{c_1 c_2}(y)$ is a gauge link, an average of the shortest
paths which connect $y_A$ and $y_B$ as products of HYP-smeared fat
links.
$\gamma_{S}$ represents the spin and $\xi_{F}$ the taste.
Here, $\chi(y_B)$ represents the staggered fermion field.
We define the Green's function as
\begin{align}
G^{f_1 f_2}_{i; c_{1} c_{2}}(x_{1}, x_{2}, y) &=\langle \chi^{f_1}_{i;
  c_{1}}(x_{1})O^{f_1 f_2}_{i}(y)\ol{\chi}^{f_2}_{i; c_{2}}(x_{2})
\rangle \,,
\end{align}
where $x_{1}$ and $x_{2}$ represents coordinates on the original
lattice with its lattice spacing $a$.
\begin{align}
x_{1}, x_{2}\in \mathbb{Z}^{4} \,, 
\qquad  y \in \mathbb{W}^{4}
\end{align}
$\mathbb{Z}^{4}$ denotes coordinate space with its lattice spacing $a$, 
and $\mathbb{W}^{4}$ denotes hypercube coordinate space with its
lattice spacing $2a$.
Now we define $\wtd{p}$ and $\wtd{q}$ as momenta defined in the
reduced Brillouin zone.
Then $p = \wtd{p} + \pi_{A}$ and $q = \wtd{q} + \pi_{B}$, where $\pi_A
= \dfrac{\pi}{a}A$.
Here, the domain of various momenta is defined as
\begin{align}
p, q \in (-\frac{\pi}{a},\frac{\pi}{a}]^{4} \,, \qquad
\wtd{p}, \wtd{q} \in (-\frac{\pi}{2a},\frac{\pi}{2a}]^{4} \,.
\end{align}
First, we apply the Fourier transformation to the Green's function $G$
as follows.
\begin{align}
\label{F}
&F^{f_1 f_2}_{i; c_{1} c_{2}}(\wtd{p} + \pi_{A}, \wtd{q} + \pi_{B}, y) 
\equiv a^{8} \sum_{x_{1}, x_{2} \in \mathbb{Z}^{4}}
e^{ i ( \wtd{p} + \pi_{A} ) x_{1} } 
e^{-i ( \wtd{q} + \pi_{B} ) x_{2} }
G^{f_1 f_2}_{i; c_{1} c_{2}}(x_{1}, x_{2}, y) \,.
\end{align}
Using the conjugate gradient algorithm, we calculate the Eq.(\ref{F})
with momentum sources of $\wtd{p}$ and $\wtd{q}$ for each gauge
configuration.
Then, we apply the Fourier transformation to $F^{f_1 f_2}_{i; c_{1}
  c_{2}}(\wtd{p} + \pi_{A}, \wtd{q} + \pi_{B}, y)$ with respect to
$y$.
After that, we make an average over the gauge configurations such that
there is no gluon field left uncontracted.
\begin{align}
H^{f_1 f_2}_{c_{1} c_{2}}(\wtd{p} + \pi_{A}, \wtd{q} + \pi_{B}, \wtd{k}) 
&\equiv \frac{1}{N} \sum_{i=1}^{N} (2a)^{4} \sum_{y \in \mathbb{W}^{4}} 
e^{-i\wtd{k} y} F^{f_1 f_2}_{i; c_{1} c_{2}} (\wtd{p} + \pi_{A}, \wtd{q} + \pi_{B}, y) \nonumber\\
& = \wtd{\delta}^{4}(\wtd{p} - \wtd{q} - \wtd{k})\wtd{H}^{f_1
  f_2}_{c_{1} c_{2}} (\wtd{p} + \pi_{A}, \wtd{q} + \pi_{B}) \,,
\end{align}
where $N$ is the number of the gauge configurations and
$\wtd{k}$ belongs to the reduced Brillouin zone.

We define
\begin{align}
\wtd{\delta}^{4}(\wtd{p}) \equiv (2a)^{4}\sum_{z \in
  \mathbb{W}^{4}}e^{i \wtd{p} z} \,.
\end{align}
Since the momentum conservation law is well respected in the reduced
Brillouin zone ($\wtd{k} = \wtd{p} - \wtd{q}$), we can rewrite $H$
as follows.
\begin{align}
\label{H}
H^{f_1 f_2}_{c_{1} c_{2}}(\wtd{p} + \pi_{A}, \wtd{q} + \pi_{B},
\wtd{k} = \wtd{p} - \wtd{q})
& =\wtd{\delta}^{4}(0)\wtd{H}^{f_1 f_2}_{c_{1} c_{2}}(\wtd{p} +
\pi_{A}, \wtd{q} + \pi_{B}) \nonumber\\
& = V \wtd{H}^{f_1 f_2}_{c_{1} c_{2}}(\wtd{p} + \pi_{A}, \wtd{q} +
\pi_{B}) \,,
\end{align}
where $V = \wtd{\delta}^{4}(0) $ is 4-dimensional volume factor.
We call $\wtd{H}$ the unamputated Green's function in this paper.

Using $\wtd{H}$ and the inverse quark propagators, 
we can obtain the amputated Green's function as follows.
\begin{align}
\wtd{\Lambda}^{f_1 f_2}_{c_{1} c_{2}}(\wtd{p} + \pi_{A}, \wtd{q} +
\pi_{B}) &=\sum_{ \substack{ C,D,\\ E,F} } \sum_{c'_1 c'_2}
   [\wtd{S}^{f_1}(\wtd{p})]_{AC; c_{1} c'_{1}}^{-1} \cdot \wtd{H}^{f_1
     f_2}_{c'_{1} c'_{2}}(\wtd{p} + \pi_{C}, \wtd{q} + \pi_{D})
   \nonumber \\
 & \hspace*{7pc} \cdot \ololSF{\gamma_{5}}{\xi_{5}}_{DF}
    [[\wtd{S}^{f_2}(\wtd{q})]^{-1}]^{\dagger}_{FE; c'_{2} c_{2}}
    \ololSF{\gamma_{5}}{\xi_{5}}_{EB} \,,
\end{align}
where $\wtd{S}(\wtd{p})$ is the quark propagator in the momentum
space \cite{Kim:2012ng}.
Let us define the projection operator $\mathbb{P}$ as follows.
\begin{align}
\mathbb{P}^{\beta}_{BA; c_{2} c_{1}} & \equiv \frac{1}{48}
\ololSF{\gamma_{S'}}{\xi_{F'}}^{\dagger}_{BA} \delta_{c_{2}c_{1}}
\\ 
\Gamma^{\alpha \beta}(\wtd{p}, \wtd{q}) & \equiv \sum_{A, B} \sum_{c_1
  c_2} [\wtd{\Lambda}^{\alpha}_{c_{1} c_{2}}(\wtd{p} + \pi_{A},
  \wtd{q} + \pi_{B}) \mathbb{P}^{\beta}_{BA; c_{2} c_{1}}] \,, \quad
\end{align}
where $\alpha$ and $\beta$ represent bilinear operators
with various spins and tastes.
Here, we call $\Gamma$ the projected amputated Green's function.
The renormalized Green's function is related to the bare one as
follows.
\begin{align}
\Gamma^{\alpha \sigma}_{R}(\wtd{p}, \wtd{q}) &= \sum_{\beta} Z_{q}^{-1}
Z^{\alpha \beta}_{O} \Gamma^{\beta \sigma}_{B}(\wtd{p}, \wtd{q}) \,.
\end{align}
Here, the subscript $R$ ($B$) denotes a renormalized (bare) quantity.
$Z_q$ is the wave function renormalization factor for the quark
fields, and $Z_O^{\alpha\beta}$ is the renormalization factor matrix
element which represents the mixing between the $\alpha$ and
$\beta$ operators.
The RI-MOM scheme prescription is 
\begin{align}
\Gamma^{\alpha \sigma}_{R}(\wtd{p}, \wtd{p}) 
= \Gamma^{\alpha \sigma}_{tree}(\wtd{p}, \wtd{p}) 
= \delta^{\alpha \sigma}\,,  
\end{align}
where $\Gamma^{\alpha \sigma}_{tree}(\wtd{p}, \wtd{p})$ is the
projected amputated Green's function at the tree level.
Therefore, the renormalization factor is obtained from the following
equation.
\begin{align}
\label{Z}
Z_{O}^{\alpha \beta} = Z_q \cdot 
[\Gamma_{B}^{-1} (\wtd{p}, \wtd{p})]^{\alpha \beta} \,.
\end{align}
\section{Numerical Results}
We use the $20^3 \times 64$ MILC asqtad lattice ($a \approx 0.12$ fm,
$am_{\ell}/am_s = 0.01/0.05$).
And we use HYP-smeared staggered fermions as valence quarks.
We perform the measurement for 5 valence quark masses (0.01, 0.02,
0.03, 0.04, 0.05), and 14 external momenta which are given in
Table~\ref{tab:momentum}.
The number of gauge configurations is 30.
We do the uncorrelated fitting and use the jackknife resampling method
to estimate statistical errors.
%
%
%
\begin{table}
\centering
\begin{tabular}{c | c | c || c | c | c || c | c | c }
\hline
\hline
$n(x, y, z, t)$ & $a|\wtd{p}|$ & GeV &
$n(x, y, z, t)$ & $a|\wtd{p}|$ & GeV &
$n(x, y, z, t)$ & $a|\wtd{p}|$ & GeV \\
\hline
$(1,0,1,3)$ & 0.5330 & 0.8835 & $(1,1,1,6)$ & 0.8019 & 1.3291 & $(2,2,2,8)$ & 1.3421 & 2.2243 \\
$(1,1,1,2)$ & 0.5785 & 0.9588 & $(1,2,1,5)$ & 0.9128 & 1.5128 & $(2,2,2,9)$ & 1.4018 & 2.3233 \\
$(1,1,1,3)$ & 0.6187 & 1.0254 & $(1,2,2,4)$ & 1.0210 & 1.6922 & $(2,3,2,7)$ & 1.4663 & 2.4302 \\ 
$(1,1,1,4)$ & 0.6710 & 1.1122 & $(2,1,2,6)$ & 1.1114 & 1.8420 & $(3,3,3,9)$ & 1.8562 & 3.0764 \\
$(1,1,1,5)$ & 0.7328 & 1.2146 & $(2,2,2,7)$ & 1.2871 & 2.1332 &             &        &        \\
\hline
\hline
\end{tabular}
\caption{
\label{tab:momentum}
The list of momenta used for our analysis.
The first column is the four vectors in the units of
$(\dfrac{2\pi}{L_s}, \dfrac{2\pi}{L_s}, \dfrac{2\pi}{L_s},
\dfrac{2\pi}{L_t})$, where $L_s$ ($L_t$) is the number of sites in the
spatial (temporal) direction.
}
\end{table}
\subsection{ Wave Function Renormalization Factor $Z_q$}
Let us consider the conserved vector current ($V_\mu \otimes S $).
The renormalization factor of the conserved currents is unity.
Therefore, we can obtain the wave function renormalization factor
$Z_q$ of the staggered quark fields from the Eq.(\ref{Z}).
\begin{align}
Z_q^{\text{RI-MOM}} = \Gamma_{B}^{\alpha \beta} (\wtd{p}, \wtd{p})\,,
\end{align}
where $\alpha = \beta = (V_{\mu} \otimes S)$. 
Here, the superscript RI-MOM denotes that the wave function
renormalization factor $Z_q$ is defined in the RI-MOM scheme.
We convert the raw data in the RI-MOM scheme into the scale-invariant
(SI) data by removing the scale-dependent part of the RG evolution
matrix as follows.
\begin{align}
Z_q^{\text{SI}} = \frac{c(\alpha_{s} (\mu_0))}{c(\alpha_{s} (\mu))} 
Z_q^{\text{RI-MOM}}(\mu), \qquad (\mu_0 = 2\text{GeV}, \quad \mu^2 =
\wtd{p}^2)
\end{align}
This Wilson coefficient $c(x)$ is calculated using four-loop anomalous
dimension given in Refs.~\cite{Aoki:2007xm,Chetyrkin:1999pq}.
In this paper, we choose $\mu_0 = 2 \text{GeV}$ to compare results
with those of other groups.

\begin{figure}[!thpb]
\subfigure[m-fit]{
\label{VxS-mass}
\includegraphics[width=0.49\textwidth]{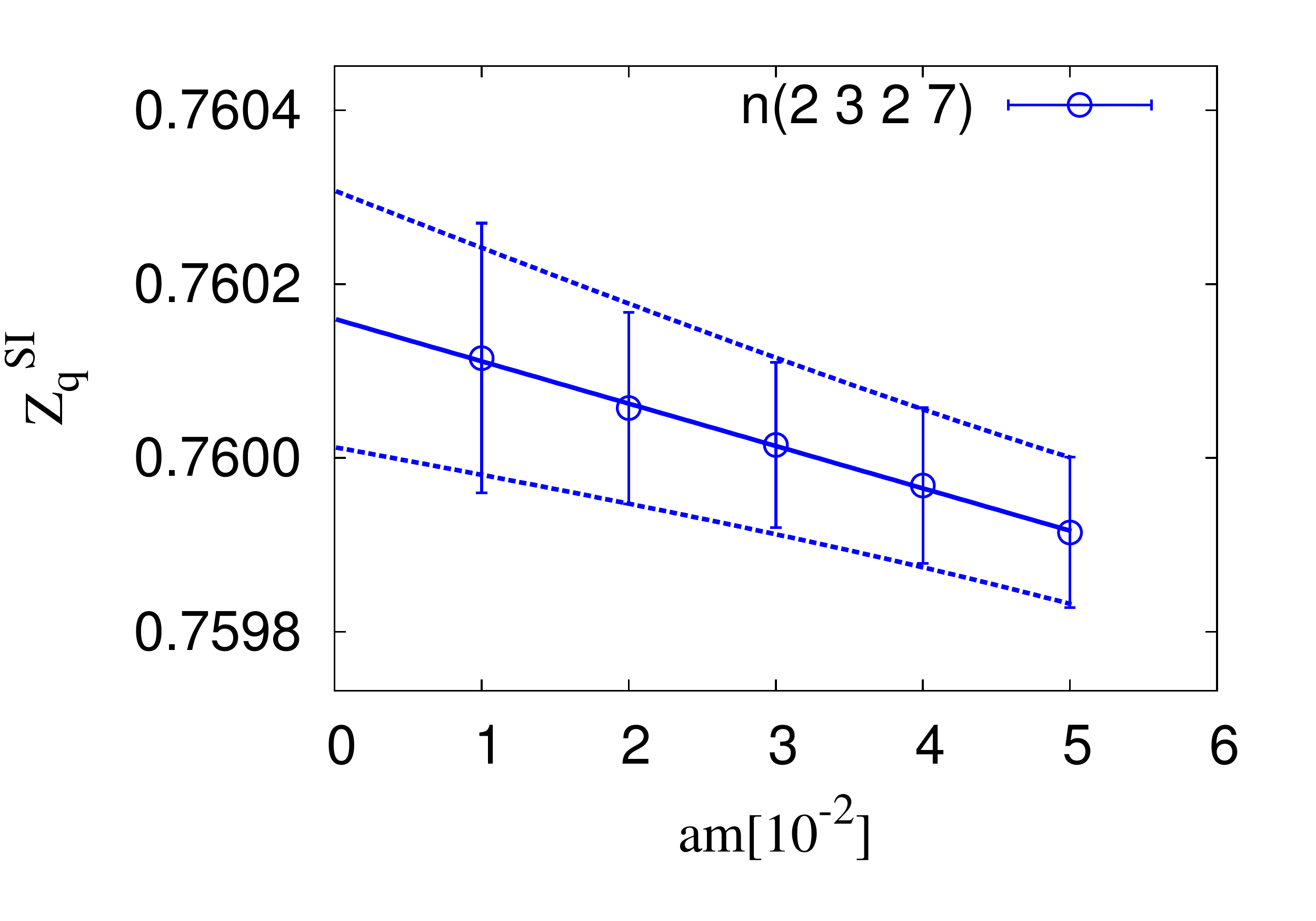}
}
\subfigure[p-fit]{
\label{VxS-mom}
\includegraphics[width=0.49\textwidth]{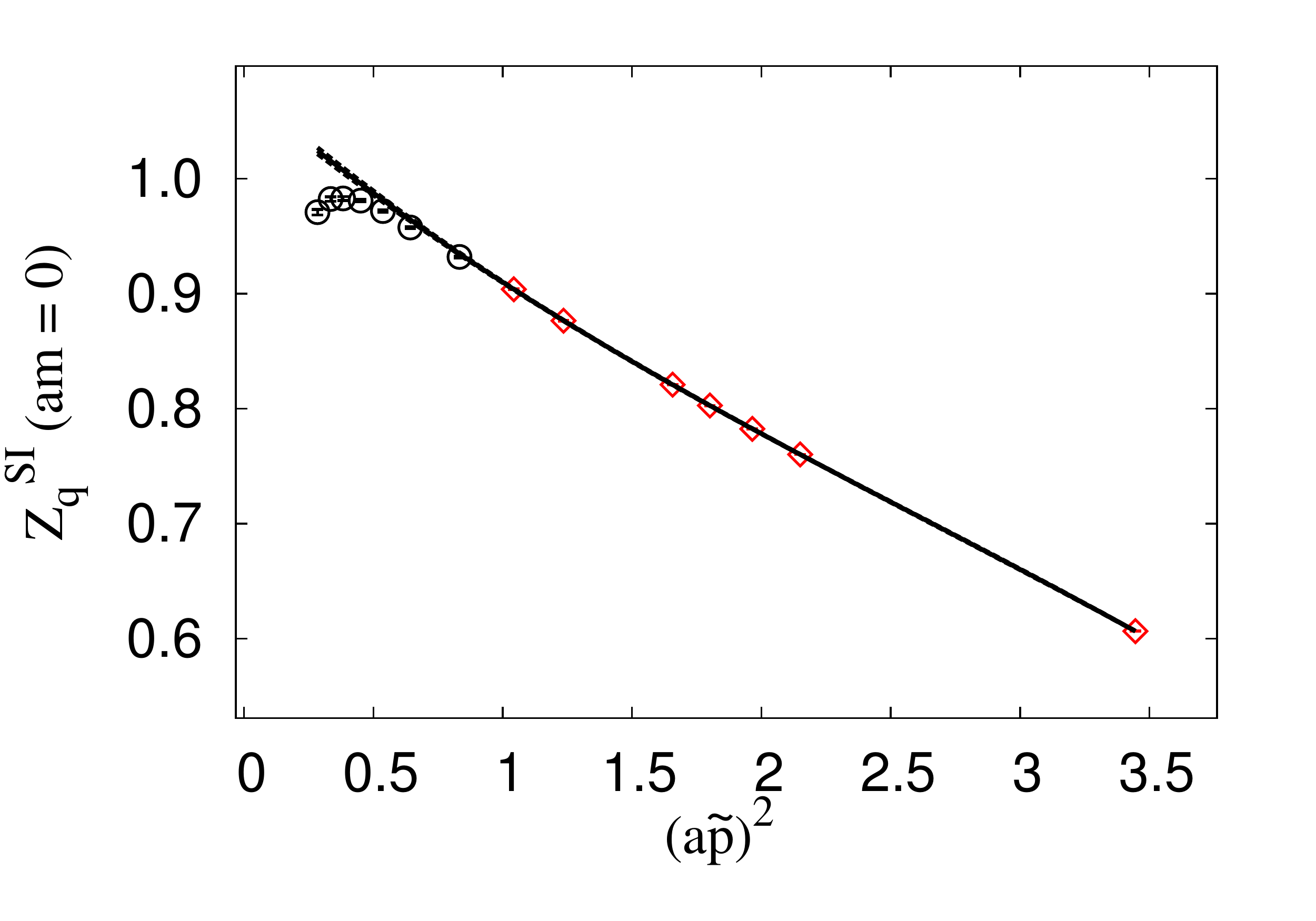}
}
\caption{ $Z_q$: The red points are within the fitting range which
  satisfies $(a\wtd{p})^2 > 1$.  }
\end{figure}

In general, the data of $Z_q$ depends on the quark mass and the 
momentum.
First, we fit the data with respect to quark mass 
for a fixed momentum to the linear function $f_{Z_q}$ as follows. 
\begin{align}
&f_{Z_q}(m, a, \wtd{p}) = a_{1} + a_{2} \cdot am \,, \qquad
\end{align}
where $a_i$ is a function of $\wtd{p}$.
We call this m-fit.
We present the m-fit results in Fig.~\ref{VxS-mass}, and the
uncorrelated fitting has $\chi^2/\text{dof} = 0.0024(62)$.
%

After the m-fit, we take the chiral limit values which corresponds to
$a_1(a,\wtd{p})$ and fit them to the following functional form.
\begin{align}
& f_{Z_q}(am=0, a\wtd{p}) = b_{1} + b_{2}(a\wtd{p})^2 +
  b_{3}((a\wtd{p})^2)^2 + b_{4}((a\wtd{p})^2)^3
\label{eq:p-fit}
\end{align}
We call this procedure p-fit.
To avoid the non-perturbative effects at small momentum region, we
choose the momentum window as $(a\wtd{p})^2 > 1$.
We present the p-fit results in Fig.~\ref{VxS-mom},
and the uncorrelated fitting has $\chi^2/\text{dof} = 0.06(16)$. 

In Eq.~\ref{eq:p-fit}, we assume that those terms of
$\mathcal{O}((a\wtd{p})^2)$ and higher order are pure lattice
artifacts.
Hence, we take the $b_1$ as the wave function renormalization factor
$Z_q$ at $\mu = 2$ GeV in the RI-MOM scheme.
We find out that $Z_q = b_1 = 1.0764(44)$, where the error is purely
statistical.

\subsection{Mass Renormalization Factor $Z_m$}
By the Ward identity, the mass renormalization factor is
\begin{align}
Z_m = \frac{1}{Z_{S \otimes S}}\,,
\end{align}
where $Z_{S \otimes S}$ is a renormalization factor of scalar bilinear
operator with scalar taste.
From the Eq.(\ref{Z}), 
\begin{align}
\left( Z_q \cdot Z_m \right)^{\text{RI-MOM}} 
=
\left( \frac{Z_q}{ Z_{S \otimes S} } \right)^{\text{RI-MOM}} 
= \Gamma_{S \otimes S} (\wtd{p}, \wtd{p}) \,,
\end{align}
where $Z_{S \otimes S} \equiv Z^{\alpha\beta}_O$ with $\alpha = \beta
= (S \otimes S)$, and $\Gamma_{S \otimes S} = \Gamma_B^{\alpha\beta}$
with $\alpha = \beta = (S \otimes S)$.
To obtain the scale-invariant(SI) quantity, we divide $(Z_q \cdot
Z_m)^{\text{RI-MOM}}$ by the RG running factors.
\begin{align}
(Z_q \cdot Z_m)^{\text{SI}} 
= \frac{c(\alpha_{s} (\mu_0) )}{c(\alpha_{s} (\mu) )} 
\cdot \frac{d(\alpha_{s} (\mu_0) )}{d(\alpha_{s} (\mu) )} 
(Z_q \cdot Z_m)^{\text{RI-MOM}} (\mu) \,, 
\quad (\mu_0 = 2\text{GeV}, \quad \mu^2 = \wtd{p}^2)
\end{align}
where $d(x)$ is the Wilson coefficient calculated using the quark mass
anomalous dimension at the four-loop level
\cite{Aoki:2007xm,Chetyrkin:1999pq}.

\begin{figure}[!thpb]
\subfigure[m-fit]{
\label{fig:SxS-mass}
\includegraphics[width=0.49\textwidth]{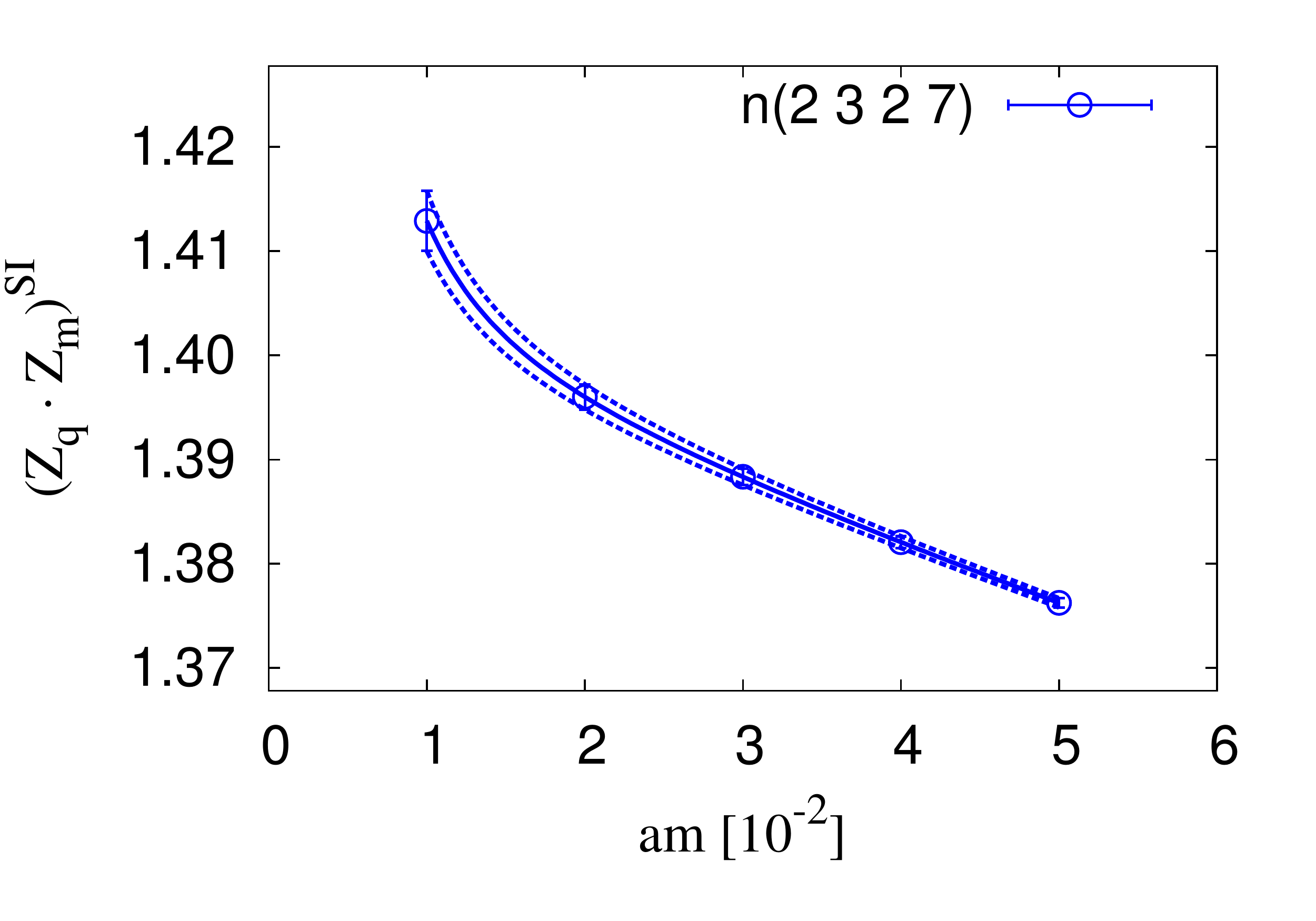}
}
\subfigure[p-fit]{
\label{fig:SxS-mom}
\includegraphics[width=0.49\textwidth]{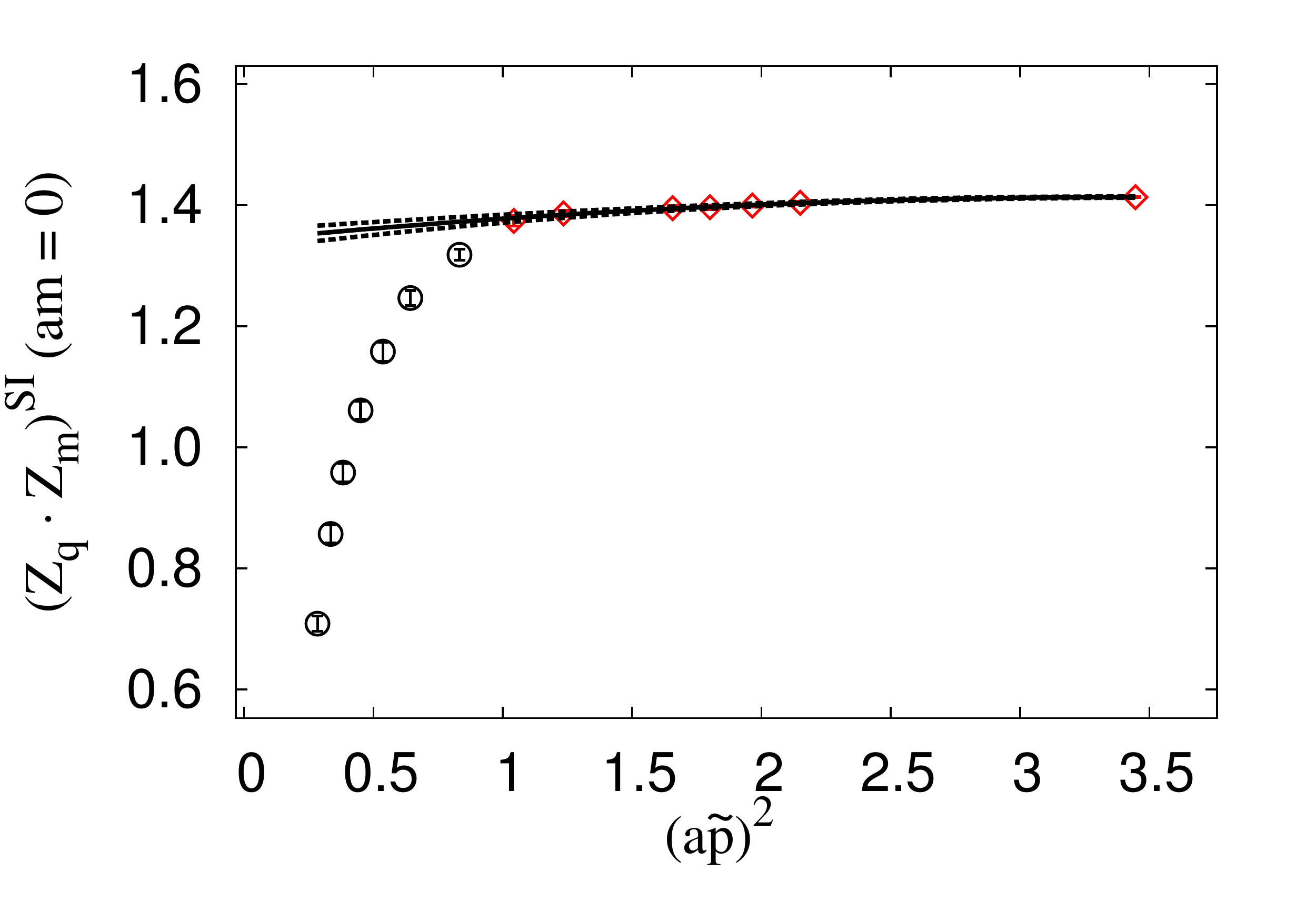}
}
\caption{ $Z_q \cdot Z_m$}
\end{figure}

In the case of m-fit, we use the following fitting function:
\begin{align}
f_{Z_q \cdot Z_m}(m, a, \wtd{p}) 
= c_{1} + c_{2}(am) + c_{3}(am)^2 + c_{4}\frac{1}{(am)^2} 
\end{align}
where $m$ is the valence quark mass.
Here, note that the $c_4$ term comes from the chiral behavior of the
chiral condensate which is proportional to $1/m^2$ due to zero modes
~\cite{Blum:2001sr}.
The sea quark determinant contributes to the chiral condensate as follows,
\begin{equation}
\langle \bar{q}q \rangle \propto \frac{(am_\ell)^2 (am_s)^1}{(am_x)^2}\,,
\end{equation}
where $m_\ell$ ($m_s$) is the light (strange) sea quark mass and
$m_x$ is the valence quark mass.
Hence, as long as we take the chiral limit of $m_\ell$ and $m_s$ at a
fixed ratio of $m_\ell/m_x = 1$, then the $c_4$ term contribution
vanishes safely.
In Fig.~\ref{fig:SxS-mass}, we show the m-fit results, and the 
uncorrelated fitting has $\chi^2/\text{dof} = 0.00008(51)$.
%

After the m-fit, we take the chiral limit values which correspond to
$c_1$.
We fit the data to the following fitting function with respect to
$(a\wtd{p})^2$.
\begin{align}
&f_{Z_q \cdot Z_m}(am=0,a\wtd{p}) 
= d_{1} + d_{2}(a\wtd{p})^2 + d_{3}((a\wtd{p})^2)^2 
\end{align}
We call this procedure p-fit.
In Fig.~\ref{fig:SxS-mom}, we present the p-fit results and
the fitting quality is $\chi^2/\text{dof} = 0.18(28)$.
Our final result is $Z_m = 1.246(15)$, where the error is purely
statistical.

\subsection{Renormalization Factors of Other Operators}
We have done the first round data analysis for the complete set of
bilinear operators.
The renormalization factor of operators($Z_O^{\alpha \alpha}$) are
calculated using Eq.(\ref{Z}) and we obtain $Z_q$ using the conserved
vector current.
Part of the preliminary results are presented in Table~\ref{tab:Z_O}. 
%

\begin{table}[!thpb]
\centering
%
%
\begin{tabular}{l l l l}
\hline
\hline
$\alpha$ & $Z_O^{\alpha \alpha}$ & (a) & (b) \\
\hline
$[S \times P]$                       &  1.079(18)  & 0.00004(23)   & 0.19(48)  \\
$[P \times A_{\mu}]$                 &  0.8947(66) & 0.00218(25)   & 0.032(74) \\
$[V_{\mu} \times V_{\mu}]$           &  0.982(11)  & 0.000003(17)  & 0.17(40)  \\
$[A_{\mu} \times A_{\nu}]$           &  1.115(27)  & 0.0000006(33) & 0.007(47) \\
$[T_{\mu\nu} \times T_{\rho\sigma}]$ &  1.293(16)  & 0.0000035(72) & 0.008(42) \\
\hline
\hline
\end{tabular}
\caption{
\label{tab:Z_O}
$Z_O^{\alpha \alpha}$ for some bilinear operators. 
Here, $\mu\ne \nu \ne \rho \ne \sigma$.
And (a) and (b) represent $\chi^2/\text{dof}$ for the m-fit
and p-fit, respectively. 
}
\end{table}
\section{Acknowledgments}
W.~Lee is supported by the Creative Research
Initiatives program (2013-003454) of the NRF grant funded by the
Korean government (MSIP).
W.~Lee acknowledges support from the KISTI supercomputing
center through the strategic support program [No. KSC-2012-G3-08].

\bibliographystyle{JHEP}
\bibliography{ref}

\end{document}